\documentclass[%
 aip,
 amsmath,amssymb,
 reprint
]{revtex4-1}

\usepackage{graphicx}
\usepackage{dcolumn}
\usepackage{bm}

\usepackage[utf8]{inputenc}
\usepackage[T1]{fontenc}
\usepackage{mathptmx}

\newcommand{\be}{\begin{equation}}
\newcommand{\ee}{\end{equation}}

\newcommand{\lf}{\left}
\newcommand{\rg}{\right}
\newcommand{\ra}{\rangle}
\newcommand{\la}{\langle}

\newcommand{\bea}{\begin{eqnarray}}
\newcommand{\eea}{\end{eqnarray}}

\newcommand{\nn}{\nonumber}

\begin{document}

\title{Two-flux tunable Aharonov-Bohm caging in a photonic lattice}

\author{V. Brosco}
\author{L. Pilozzi}%
 \email{laura.pilozzi@isc.cnr.it}
\affiliation{ 
Institute for Complex Systems, National Research Council, Via dei Taurini 19, 00185 Rome, Italy 
}%

\author{C. Conti}
\affiliation{ 
Institute for Complex Systems, National Research Council, Via dei Taurini 19, 00185 Rome, Italy 
}%
\affiliation{Department of Physics, University Sapienza, Piazzale Aldo Moro 5, 00185 Rome, Italy}

\date{\today}

\begin{abstract}
We study the  Aharonov-Bohm caging effect in a one-dimensional lattice of theta-shaped units defining a chain of interconnected plaquettes, each one threaded by two synthetic flux lines. 
In the proposed system, light trapping results from the destructive interference of waves propagating along three arms, this implies that  the caging effect is tunable and it can be controlled by changing the tunnel couplings $J$. These features reflect on the diffraction pattern allowing to establish a clear connection between the lattice topology and the resulting AB interference.

\end{abstract}

\maketitle

\section{Introduction}
In the early 1980s the geometrical interpretation of some phenomenological observables has introduced a new paradigm for the explanation of different effects\cite{Vanderbilt2018}, modifying, for example, our view of non-relativistic quantum phenomena such as the quantum Hall effect \cite{Thouless1982}  and prompting new developments and discoveries of paramount relevance ranging from modern polarization theory\cite{Resta1993} to topological phases\cite{hasan2010}. 

Nowadays, the geometry of the Hilbert space, with its metric, defining the distance between two quantum states, and its connection\cite{Wilczek1989}, fixing the phase accumulated along quantum trajectories, is a central object in condensed matter research, holding great promises for quantum information applications.
In photonics and atomic physics the quest to engineer and control  the geometric and topological properties of artificial lattices fostered remarkable efforts to 
implement effective electromagnetic fields for neutral particles.\cite{aidelsburger2018} Just to mention a few examples, uniform magnetic fields were achieved in optical lattice-based experiments,\cite{Aidelsburger2011} in ring resonator arrays \cite{hafezi2013}, in optomechanical systems \cite{schmidt2015}.
In photonic lattices, artificial gauge fields were generated using different techniques, {\sl i.e.} introducing topological defects in two-dimensonal structures,\cite{rechtsman2012, lumer2019} applying time-dependent modulation\cite{fang2012,goldman2014,joerg2017}, employing synthetic modal dimensions\cite{lustig2019} and, very recently, controlling the orbital angular momentum of the input light beam.\cite{joerg2020}
What underlies most of the observations carried out in the above systems is the first discovered and most basic consequence of the existence of gauge-fields, the Aharonov-Bohm (AB) effect \cite{aharonov1959}. 
The paramount importance of this effect ranges from metrological applications to basic physics\cite{batelaan2009}. It is a non-local effect, arising from the interference of electron beams traveling along paths enclosing a magnetic flux.
As first recognized by Wu and Yang \cite{wu1975}, it naturally leads to  the concept of  path-dependent
phase factor as a basis to describe electromagnetism and gauge theories in general. 
Furthermore, as it clearly emerges in path-integral derivations, AB interference reflects the multiply connected nature of the space and it may have impressive consequences on transport. An example is Aharonov-Bohm caging a single-particle localization effect arising from the interplay between the lattice structure and the magnetic flux, first predicted by J. Vidal et al. \cite{vidal1998}
for two-dimensional electronic lattices and subsequently extended and experimentally verified in different contexts, \cite{leykam2018,abilio1999,rizzi2006}  including photonic lattices, see {\sl e.g.} Refs. \onlinecite{fang2012,mukherjee2018,kremer2020}.

\begin{figure}[t]
	\includegraphics[width=0.6\columnwidth]{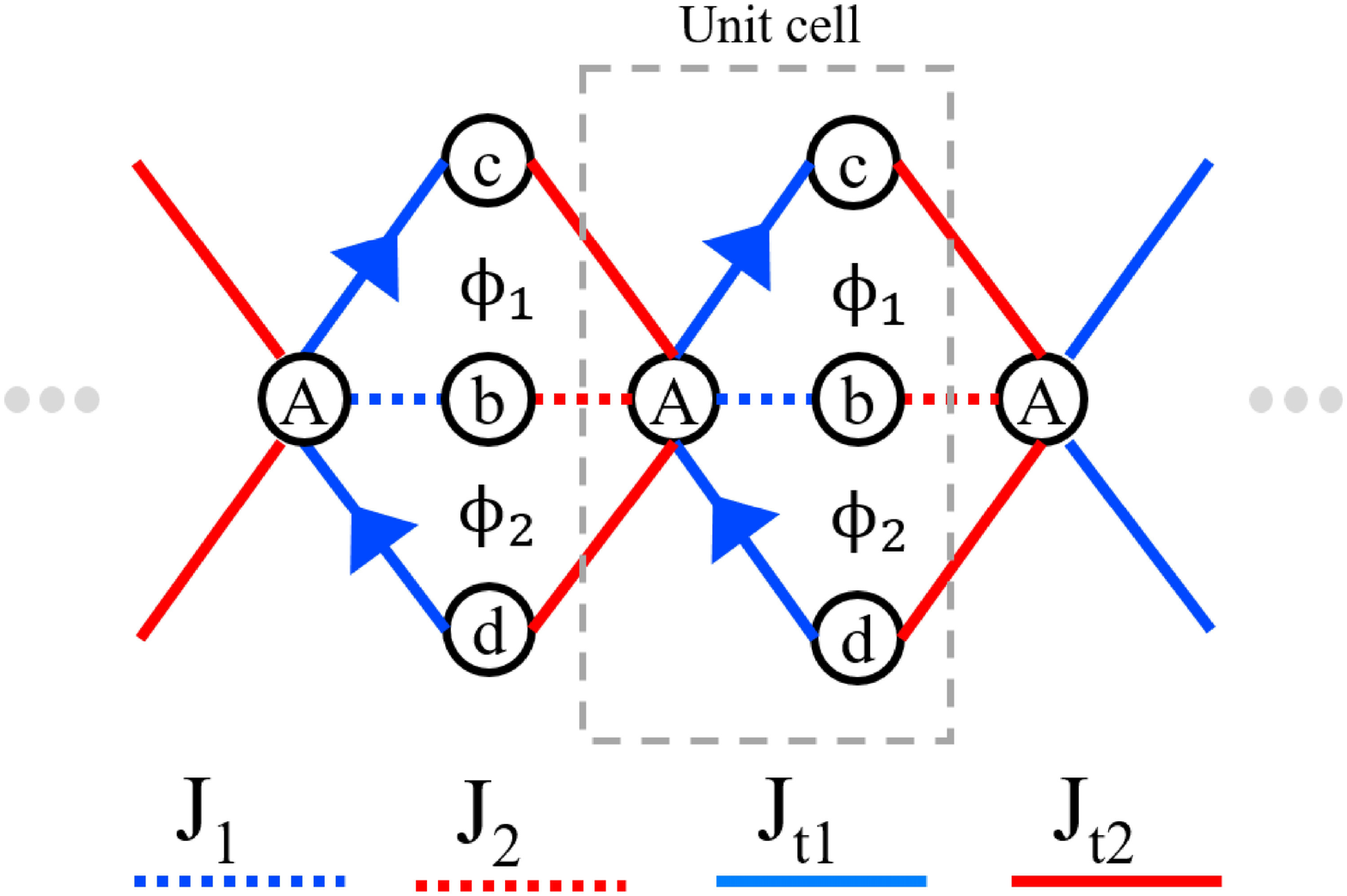}
	\caption{Sketch of the structure of the $\theta$-lattice composed of interconnected plaquettes, each one threaded by two synthetic flux lines, $\phi_1$ and $\phi_2$. An arrow
		means a phase $e^{i\phi}$. } \label{AB-rings1}
\end{figure}

 In the present work we study the propagation of light through a one-dimensional array of theta-shaped plaquettes  threaded by two fluxes as shown in Fig. \ref{AB-rings1}, that, for brevity,  we call $\theta$-lattice.
 The presence of two fluxes allows us to investigate in a simple but non-trivial framework the signatures of Aharonov-Bohm interference on diffraction patterns highlighting its topological significance and showing how in this case the caging effect becomes fully tunable.

\section{Model}

We consider a one-dimensional lattice of theta-shaped units as shown in Fig.\ref{AB-rings1}.
Its unit cell consists of four sites indicated respectively as $A$, $b$, $c$ and $d$. The three arms of each ring, defined by the sites $A_n$, $A_{n+1}$ and respectively $b_n$, $c_n$ and $d_n$, enclose two synthetic flux lines, indicated respectively as $\phi_1$ and $\phi_2$. 
\\
The Hamiltonian of the $\theta$-lattice can thus be written as: 
\bea
H&=&\sum_n \lf[J_{\rm t1} \lf(a^\dag_{n}c_n e^{-i\phi_1}+a^\dag_{n}d_n e^{i\phi_2}\rg)+J_{\rm t2}\lf( a^\dag_{n}d_{n-1}+a^\dag_{n}c_{n-1}\rg)+\rg.\nn\\
&& \lf.+J_1a^\dag_{n}b_n + J_2 a^\dag_{n}b_{n-1}+ {\rm  H. c.} \rg] \label{re}
\eea
where $m_n$, $m^\dag_n$  with $m= a,\, b,\, c,\, d$ are bosonic annihilation and creation operators corresponding to the sites $A,\, b,\, c,\, d$ of the cell $n$. 
Switching to $k$-space we obtain: 
\bea
\mathcal{H}(k,\phi_1,\phi_2) &=& \sum_k \big[ J_b(k)\, a^\dag_k b_k +  J_c (k,\phi_1)\,  a^\dag_k c_k+\nn\\ & &  J_d (k,\phi_2)\, a^\dag_k d_k+
{\rm H.c.} \big] \label{mom}
\eea
\noindent
where $m_k=\frac{1}{\sqrt{N}}\sum_n m_n e^{ikn}$, with $N$ denoting the number of unit cells in the lattice, while $J_b(k)=J_1 +J_2 e^{-i k} $ and $J_c(k,\phi)=J_d(k,-\phi)=J_{t1}e^{-i\phi} +J_{t2} e^{-i k}$, where $J_1$  and  $J_{t1}$ and $J_2$ and $J_{t2}$ denote the intra- and inter-cell hopping amplitudes. The $\phi_1$, $\phi_2$ dependence of $J_c$ and $J_d$ is due to the presence of the synthetic  gauge fields.
We note that setting $J_{t2}=J_{t1}=0$ the $\theta$-lattice reduces to the non-abelian Lieb lattice model\cite{brosco2020} while setting $J_b=0$  it reduces to the standard rhombi chain with a flux $\phi_T=\phi_1+\phi_2$.  
  
The Hamiltonian $\mathcal{H}(k,\phi_1,\phi_2)$ is invariant, up to a gauge transformation, under permutations of the three arms b, c, and d, \textit{i.e.} under elements of the non-Abelian group $S_3$. This implies that the state
\be
|\phi_{s}(k)\ra=J_b |b_k\ra+J_c |c_k\ra+J_d |d_k\ra,
\ee
invariant under elements of $S_3$, yields two dispersive modes:
\be
|\psi_{\pm}(k)\ra=\frac{1}{\sqrt{2}}\lf [|a_k\ra \pm \frac{|\phi_{s}(k)\ra}{ \Delta(k,\phi_1,\phi_2)}\rg]
\ee
with longitudinal momenta $ \kappa_{\pm}(k,\phi_1,\phi_2)=\pm\Delta(k,\phi_1,\phi_2)$ and  
 \be
 \Delta(k,\phi_1,\phi_2)=\sqrt{|J_b(k)|^2+|J_c(k,\phi_1)|^2+|J_d(k,\phi_2)|^2} .
 \ee

On the other hand, the states 
\begin{eqnarray}
|w_1\ra=J_b^*|c_k\ra-J_c^*|b_k\ra \label{deg1} \\
|w_2\ra=J_b^*|d_k\ra-J_d^*|b_k\ra 
\label{deg2}
\end{eqnarray}
spanning a two-dimensional non-invariant subspace of $S_3$ must be degenerate for all $J$'s.  These states thus yields two non-dispersive modes for any $\phi_1$ and $\phi_2$ with longitudinal momentum $\kappa=0$. The presence of these modes  underlies an SU(2) non-abelian gauge symmetry.

\begin{figure}[t!]
\includegraphics[width=1\columnwidth]{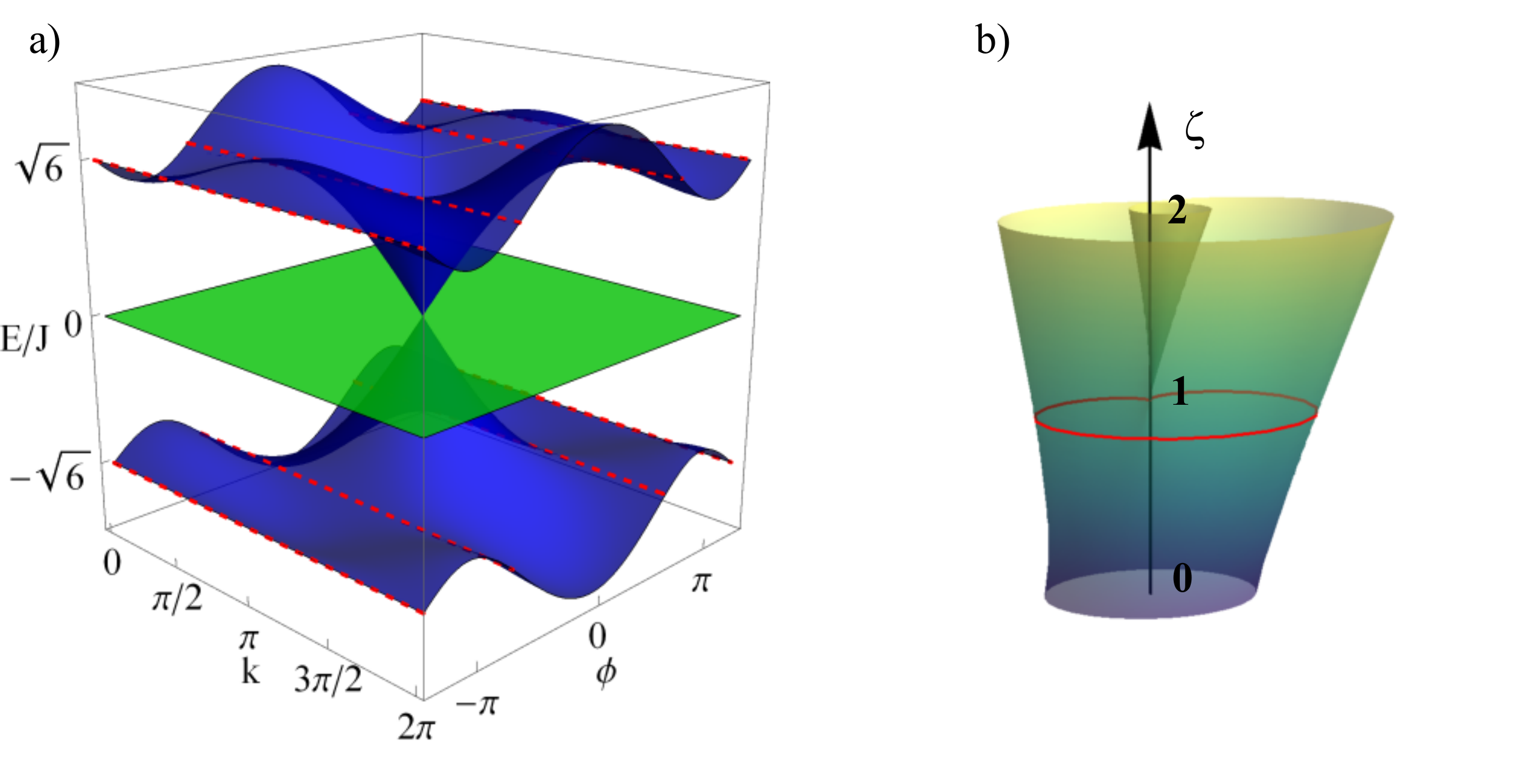}
\caption{a) Spectrum of $\mathcal{H}(k,\phi_1,\phi_2)$ for $J_i=J$ with i $\in (1,2,t1,t2)$ and $\phi_1=\phi_2=\phi$. The red dashed lines show the $\phi$ values that give four flat bands. b) Cylindrical plot of the surface  $\rho(\phi, \zeta)=1+\zeta \cos(\phi)$ } \label{AB-rings}
	\end{figure}

The overall structure of the spectrum for $\phi_1=\phi_2=\phi$  with $J_i=J$ for $i=(1,\,2,\,t1,\,t2)$ can be seen in Fig.\ref{AB-rings}a) showing a band crossing at ($k,\phi$)=($\pi$,0) and gaps for $\phi\ne 0$.  We notice that for certain values of the coupling $J$ and flux per plaquette \textit{all} bands in the energy dispersion became flat as indicated by the red dashed lines. This condition, a result of a destructive interference induced by the synthetic magnetic field, gives rise to light trapping and corresponds to the AB caging effect.

 \begin{figure*}[t]
\includegraphics[width=1.5\columnwidth]{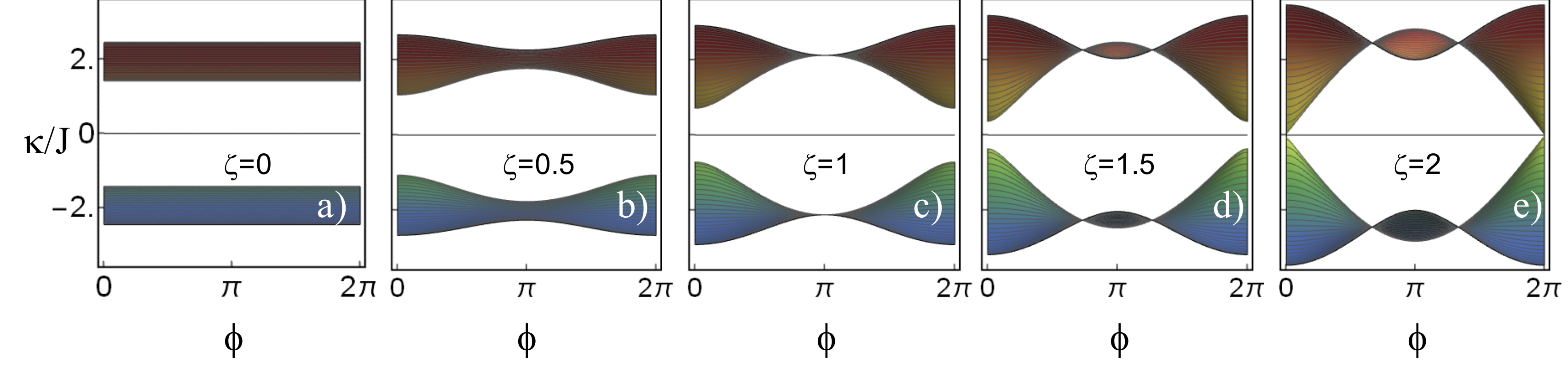}	\caption{Spectrum support on the $(\kappa,\phi)$ plane for different values of $\zeta$.} \label{fig-spectrum-suppotrt}
 \end{figure*}

\section{Aharonov-Bohm caging} 
A peculiarity of the two-flux model is that the caging arises due to the destructive interference of waves propagating along three arms. This implies that, at variance with the standard two-arm single-flux AB cages\cite{Longhi2014},  the values of $\phi_1$ and $\phi_2$ where the caging effect appears can be controlled by changing the tunnel couplings J. 
We remark that when caging arises due to the destructive interference of waves propagating along two arms it has to be necessarily located at $\phi=\pi$, {\sl i.e.} the total amplitude is given by the sum of two identical terms having opposite sign. On the contrary, as stated above, in the $\theta$-lattice we find different caging conditions depending on the tunnel couplings.
In particular, when all J's are equal,  caging appears for $\phi_1=\phi_2=2\pi/3$  and $\phi_1=\phi_2=4\pi/3$ reflecting the trigonal symmetry of the unit cell. 
For arbitrary values of the $J$'s  and $\phi_1=\phi_2=\phi$ the  condition to have dispersionless bands  can be written as  follows
\be 1+ \zeta \cos(\phi)=0 \label{eq-cage}\ee 
 where $\zeta=2J_{t2}J_{t1}/(J_1 J_2)$.
In Fig.\ref{AB-rings}b) we show a cylindrical plot of the surface  $\rho(\phi, \zeta)=1+\zeta \cos(\phi)$, where we clearly distinguish three cases:
for $\zeta>1$ we have two values of $\phi \in[0,2\pi\,[$ where Eq.\eqref{eq-cage} is satisfied, for $\zeta<1$ the caging condition is never fulfilled, while  for $\zeta=1$ Eq.\eqref{eq-cage}  admits only the solution $\phi=\pi$ as in the case of the standard two-arm AB caging.

\begin{figure}[b]
	\includegraphics[width=\columnwidth]{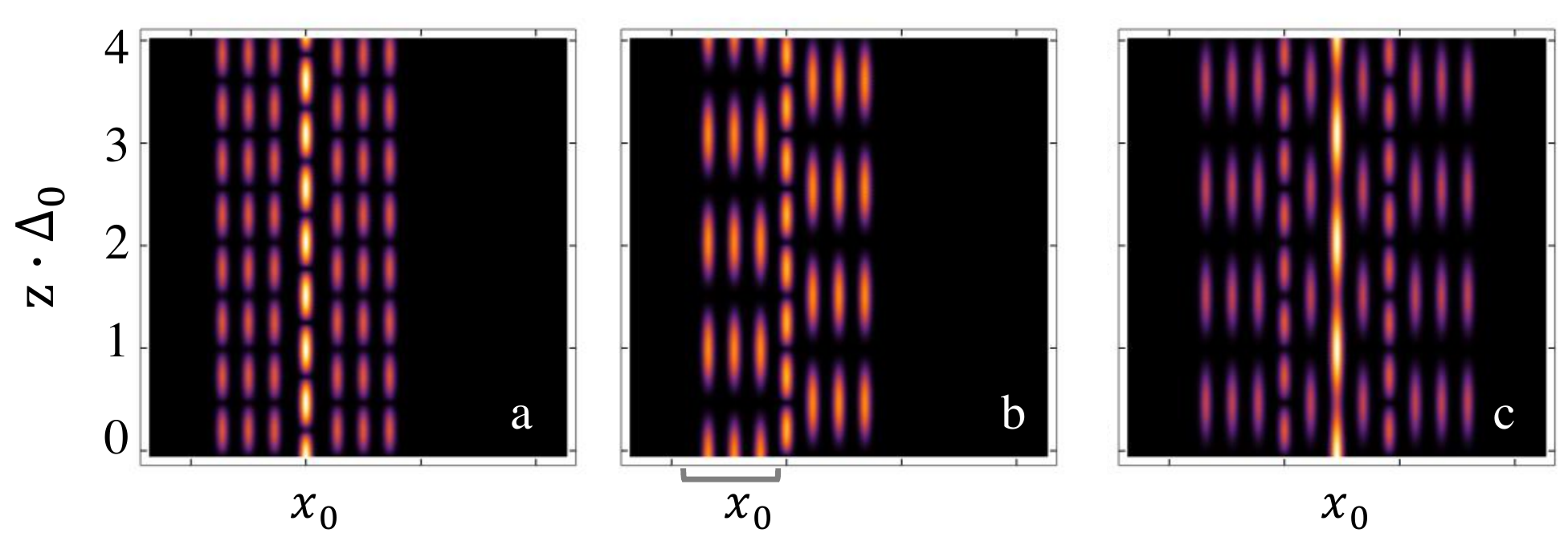}
	\caption{Light dynamics in the presence of AB caging for three different injection configurations: (a) $|\psi_0\ra=| a_{n_0}\ra$, (b) $|\psi_0\ra=\frac{1}{\sqrt{3}}(| b_{n_0}\ra+| c_{n_0}\ra+| d_{n_0}\ra)$  and (c) $|\psi_0\ra= |c_{n_0}\ra$. Lattice parameters: $J_{1}=J_{2}=J_{t1}=J_{t2}=J$, $\phi_1=\phi_2=\phi=2\pi/3$, $\Delta_0=\sqrt{6}J$, $N=40$, $n_0=20$. }\label{sim1}
\end{figure}

 To further analyze how caging arises when $\phi_1=\phi_2=\phi$, in Fig.\ref{fig-spectrum-suppotrt} (a-e) we show the evolution of the 
 quasi-energy spectrum support as $\zeta$ increases from $\zeta=0$ to $\zeta=2$. 
At $\zeta=0$, corresponding to $J_{t1} J_{t2}=0$,  the spectrum is clearly  $\phi$-independent; as we increase $\zeta$ we find a pseudo-localization region around $\phi=\pi$ that evolves in a fully localized spectrum for $\zeta=1$; eventually for $\zeta>1$, the spectrum support shows two nodes, signaling the emergence of genus 2 AB caging.

\begin{figure}[b]
	\includegraphics[width=\columnwidth]{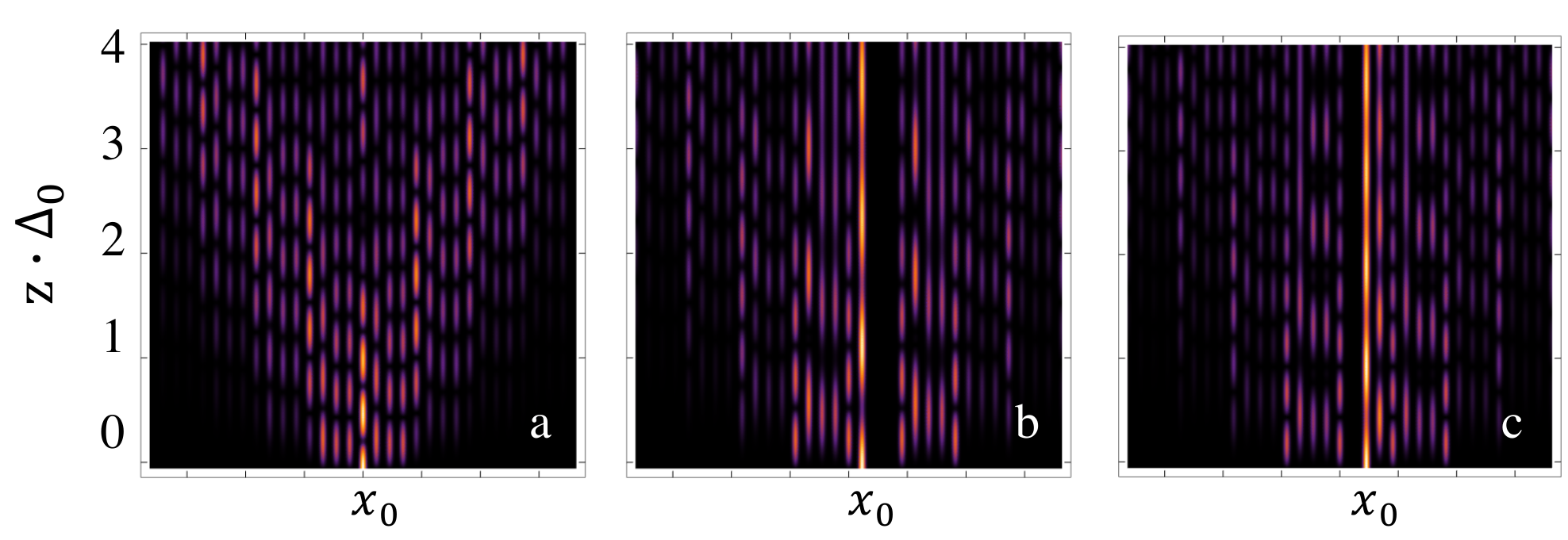}
 \caption{Dispersive light dynamics for  $\phi_1=\phi_2=\phi=\pi$, {\sl i.e.} away from the AB caging condition, for three different injection configurations: (a) $|\psi_0\ra=| a_{n_0}\ra$, (b) $|\psi_0\ra=| b_{n_0}\ra$  and (c) $|\psi_0\ra=| c_{n_0}\ra$.  Other lattice parameters as in Fig. \ref{sim1}. } \label{sim2}
\end{figure}

Let us now consider the light dynamics in the different caging regimes. As discussed by several Authors, see \textit{e.g.} \cite{Longhi2014,christodoulides2003}, assuming evanescent coupling of single-mode waveguides, it is described by the following coupled mode equations:

\be\lf\{
\begin{array}{rcl}
i\partial_z a_n&=&  J_{1} b_{n}+ J_{2} b_{n-1}+J_{t1}( e^{i \phi_1}c_n+e^{-i \phi_2}d_n)+\\
& & + J_{t2}( c_{n-1}+d_{n-1})\\ [0.1cm]
i\partial_z b_n&=&  J_{1} a_{n}+ J_{2} a_{n+1}\\[0.1cm]
i\partial_z c_n&=&  J_{t1}e^{-i \phi_1}a_n+ J_{t2}a_{n+1}\\[0.1cm]
i\partial_z d_n&=&  J_{t1}e^{i \phi_2}a_n+ J_{t2}a_{n+1}
\end{array}\rg.\ee
where $\partial_z $ indicate the partial derivative with respect to $z$.
Solving numerically the above equations on a finite lattice with $N$ unit cells (4$N$ sites) and open boundary conditions yields the results shown in Figs. \ref{sim1} and \ref{sim2}.

In Figure  \ref{sim1} we simulate the propagation of a light beam injected at $z=0$ in a $\theta$-lattice consisting of $N$ unit cells, $4N$ waveguides, with homogeneous tunnel couplings $J_i=J$ and fluxes $\phi_1=\phi_2=\phi=2\pi/3$.
For these parameters the dispersive bands $\kappa_{\pm}$ become flat and, independently of the precise position and energy of the incoming beam,  light gets trapped on a cluster of few waveguides. Only the structure of the caging cluster depends on the initial condition. This is due to the fact that,  depending on the initial condition, different localized bands enter the dynamics. When the light is injected in a site $A$ only the upper and lower bands are dynamically occupied; caging then implies that only the waveguide $A_n$ and  the six surrounding waveguides $b_n,c_n,d_n$ and $b_{n-1},c_{n-1},d_{n-1}$ are populated as shown in Fig.\ref{sim1}a). 
The wavelength, $\lambda_0$, of the oscillations between the upper and lower bands is clearly given by the inverse of the spectral gap {\sl i.e.} $\lambda_0=1/(2\Delta_0)$ with $\Delta_0=\Delta(k, 2\pi/3,2\pi/3)=\sqrt{6}J$.
A somewhat similar situation arises when light is injected symmetrically in the waveguides  $b_n,c_n,d_n$, {\sl i.e.} creating the initial state  $|\psi_0\ra=\frac{1}{\sqrt{3}}(| b_{n_0}\ra+|c_{n_0}\ra+|d_{n_0}\ra)$. The peculiar structure of the initial state implies that  in this case, shown in Fig. \ref{sim1}(b), the light beam undergoes  oscillations between the cell $n$ and $n+1$ along $z$ without modifying its shape. Eventually in Fig. \ref{sim1}(c) we show the propagation of a light beam injected from the site $c_n$. In this case the evolution involves also the degenerate bands and the signal spreads over three unit cells.

When the caging condition is not fulfilled, light spreads to the entire lattice. 
This situation is considered in Fig.\ref{sim2} where we set $\phi_1=\phi_2=\pi$ and the other parameters as in Fig. \ref{sim1}. These values of the fluxes are special under many respects: first, as discussed in the following section, they yield,  Fig.\ref{sim2}a), a weaker dispersion as compared {\sl e.g.} to the case  $\phi_1=\phi_2=0$, and second, they yield  destructive Aharonov-Bohm interference on specific sites of the array. For example, as shown in Fig.\ref{sim2}(b), for  a "$b$" type injection waveguide, the propagation does not involve the sites "$c$" and "$d$" in the same plaquette, while,  as shown in Fig.\ref{sim2}(c), for  a "$c$" type injection waveguide, the propagation does not involve the site "$b$" in the same plaquette.
This is due to the fact that, at $\phi_1=\phi_2=\pi$ there are two tunneling paths from $b$ to $c$ and from $b$ to $d$ having opposite signs and equal amplitudes.

\section{Signatures of Aharonov-Bohm interference in the diffraction patterns}
\begin{figure}[t]
	\includegraphics[width=0.45\columnwidth]{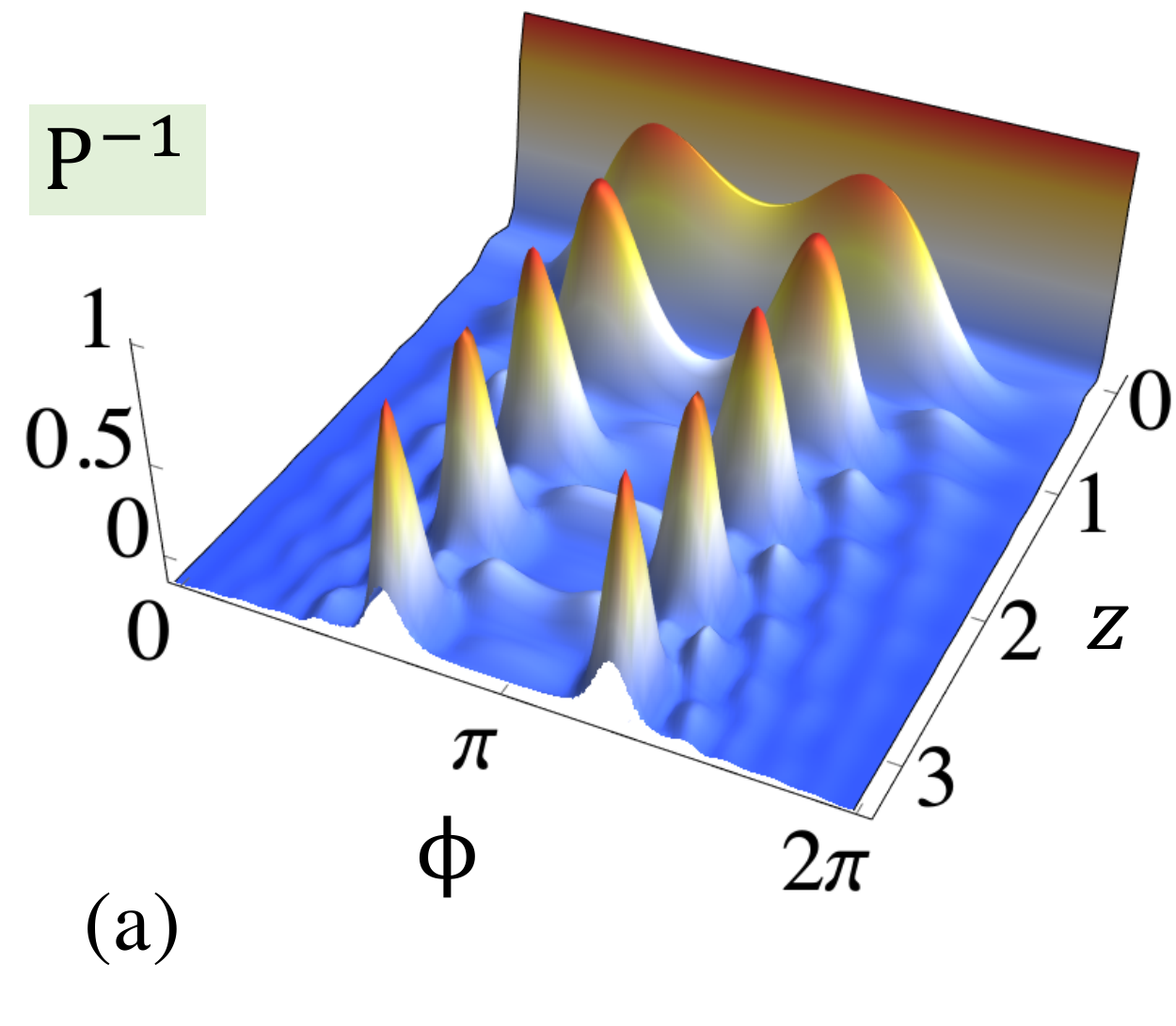} \hspace{0.2 cm}
	\includegraphics[width=0.45\columnwidth]{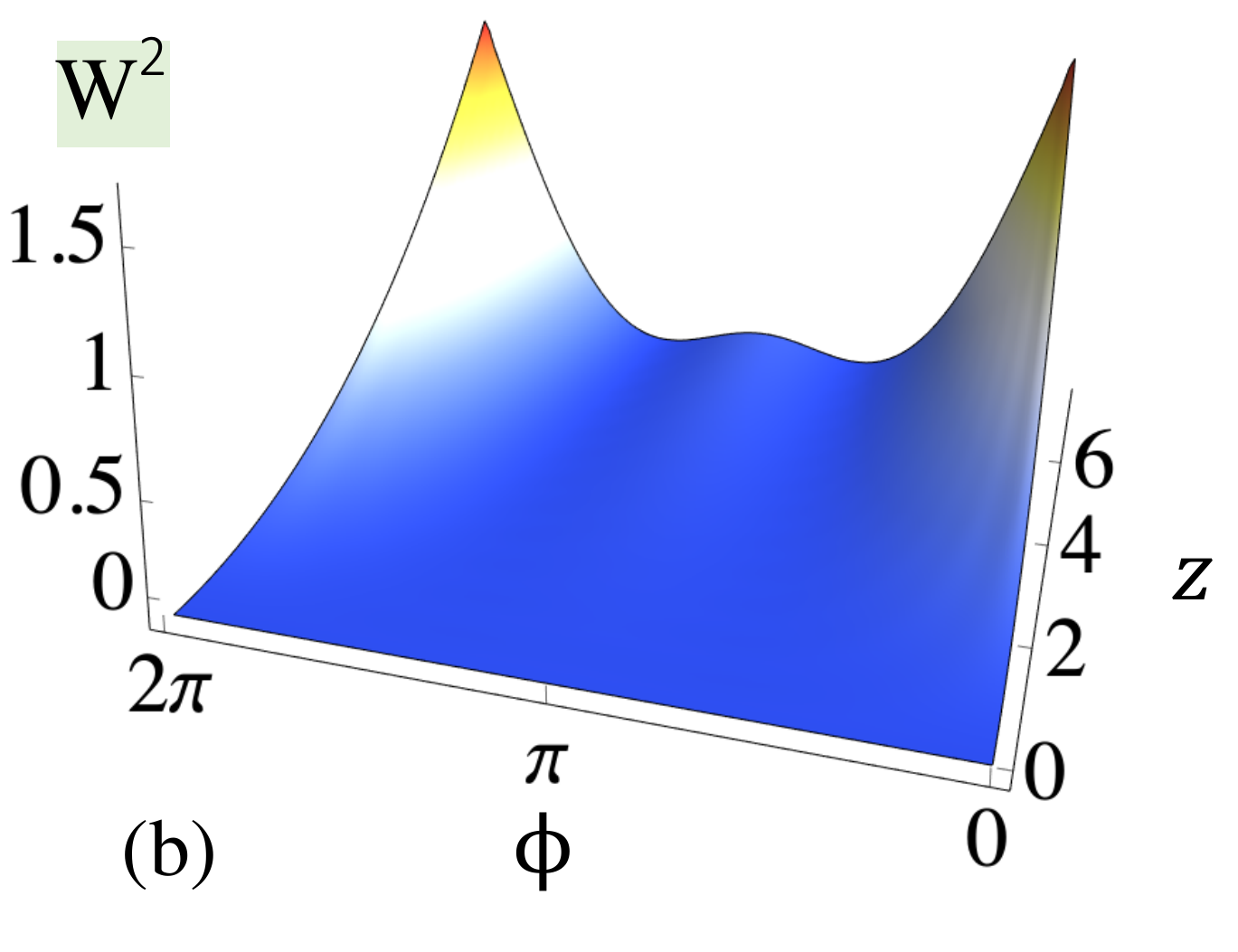}
	\caption{Inverse participation ratio (a) and average width (b) as functions of $z$ and $\phi$. Other lattice parameters as in Fig. \ref{sim1}} \label{diffraction}
\end{figure}
\begin{figure}[t!]
	\includegraphics[width=0.9\columnwidth]{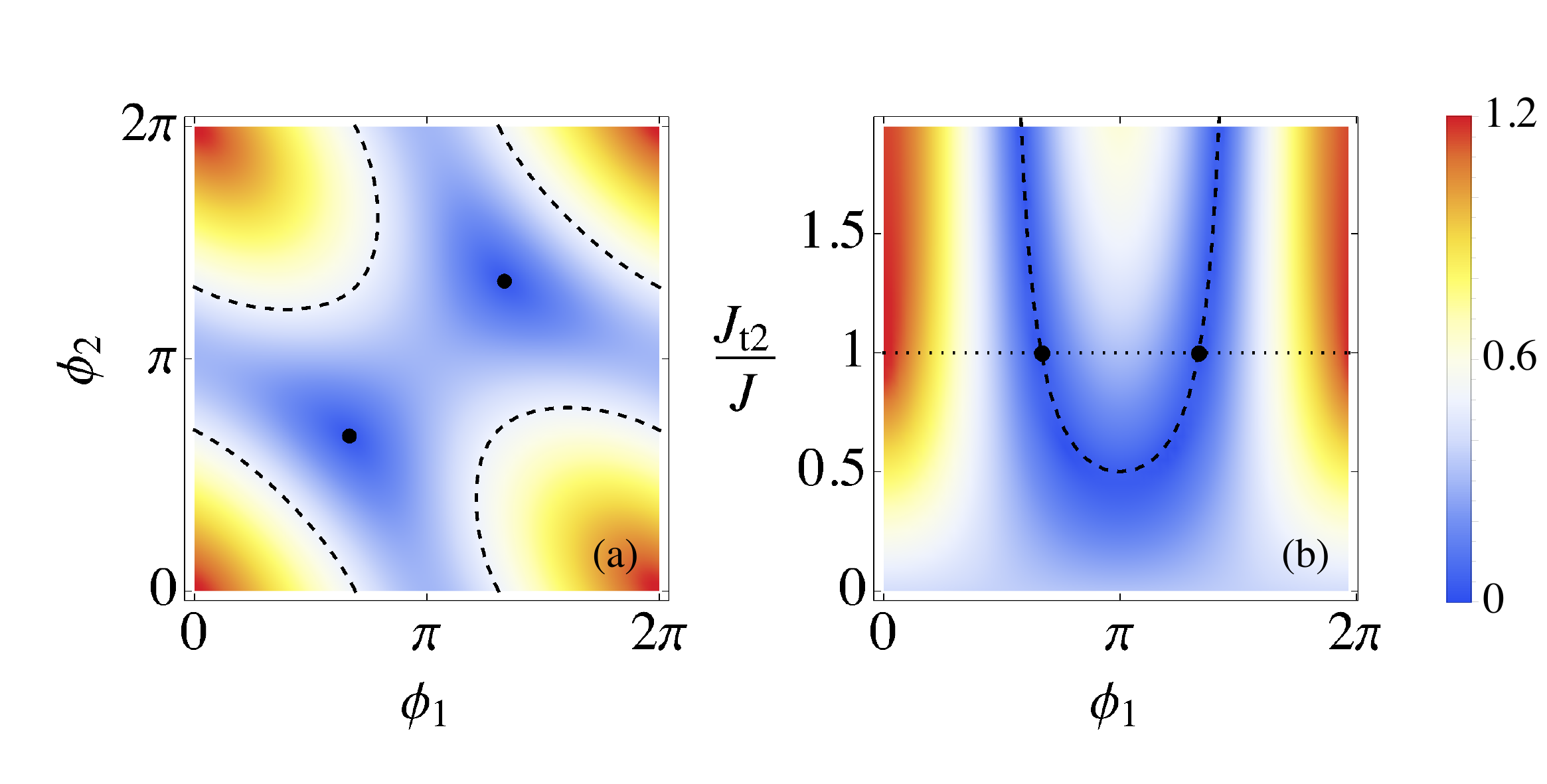}  
	\caption{Density plot of the width $W$ as a function of $\phi_1$ and $\phi_2$ and as a function of $J_{t2}$ and $\phi=\phi_1=\phi_2$.} \label{phase-diagram}
\end{figure}

 Beside inducing AB caging, synthetic gauge fields modulate light propagation in photonic lattices through AB interference, mimicking the action of their electromagnetic counterparts   and yielding  synthetic-flux dependent diffraction effects.
Purpose of the present section is to highlight how  these effects arise in the $\theta$-lattice. 
To characterize diffraction for different values of the synthetic fluxes we will focus on two quantities, namely,
the inverse participation number, $P^{-1}$, defined as 
\be
P^{-1}=\sum_{x}|\Psi_x|^4,
\ee
 where $\Psi_x$ denotes the field's amplitude at position $x=na$ along the lattice, with "a" the lattice period, and the average square width, $W^2$, defined as
\be \label{eq:w}
W^2=\frac{\la x^2 \ra-\la x \ra^2}{N^2}
\ee
with $\la x^\alpha \ra=\sum_x x^\alpha |\psi_x|^2$.
The inverse participation number is always smaller or equal 1 and it gives a measure of the number of sites where photons are confined, specifically we have $P= 1$ when light is confined to a single waveguide and $P\sim m$ when light is confined to a cluster of $m$ waveguides. The average width $W$ is useful to characterize how the signal disperse, it equals zero in the presence of caging and in standard photonic waveguide lattices it grows as $z^2$.
In Figure \ref{diffraction}(a) we plot the participation ratio as a function of $z$ and $\phi$ for a lattice with  homogeneous tunnel couplings and fluxes, {\sl i.e.} $\phi_i=\phi$ and $J_i=J_{ti}=J$ with $i=1,2$. We assume that the system is initially prepared in the fully localized state $|a_{n_0}\ra$, at $z=0$ we thus have $P^{-1}=1$ independently of $\phi$. As $z$ increases, light start dispersing and we clearly see the emergence of two peaks at $\phi=2\pi/3$ and $\phi=4\pi/3$ due to AB caging. We also notice that $P^{-1}$ has a strongly oscillating behavior with $z$ that is associated with dynamic oscillations between different bands.
The presence of these oscillations may hinder the characterization of the difference interference regimes by simply measuring the amplitude of the fields in a small cluster of sites for a given propagation length $z=\bar z$. For this purpose, the square width $W^2$ defined in Eq.\eqref{eq:w} may be more appropriate as  we show in Figure \ref{diffraction}(b). There, we notice in particular the emergence of a smooth double-well structure associated with AB interference.  Having a monotonic behavior as a function of $z$, $W$ can be used to characterize the different diffraction regimes for different values of the synthetic fluxes and $J$'s. This is what we do in Fig. \ref{phase-diagram} (a-b) to illustrate the tunability of the caging effect in the $\theta$-lattice. 
 In Fig. \ref{phase-diagram} (a) we show a density plot of the width $W$ calculated at $z J=10$ for the system initially prepared in the state $|a_{n_0}\ra$, as a function of $\phi_1$ and $\phi_2$  for homogeneous tunnelings. We clearly see that the contour $W=0.5$ represented by the dashed black line essentially allows us to distinguish between a weakly dispersing region including the caging points $\phi_1=\phi_2=2\pi/3$ and $\phi_1=\phi_2=4\pi/3$ and  a strongly dispersing region for $\phi \lesssim \pi/2$.
 In  Fig. \ref{phase-diagram} (b) we plot $W_{z=10 J}$ as a function of   $\phi$ and $J_{t2}$  setting all other tunnel couplings to $J$.
In this figure the black dashed line indicates the caging condition given by Eq. \eqref{eq-cage}. We notice that when $J_{t2}$ becomes much larger than $J$ the tunability essentially disappears, this is due to the fact that increasing $J_{t2}$ corresponds to decrease the weight of interference paths going through the site $b$ bringing the lattice back to the single flux regime.

\section{Conclusions}
We presented a theoretical study of transport of light in a strip of theta-shaped plaquettes subjected to synthetic magnetic fields. We showed how to realize Aharonov-Bohm cages that prevent the photon beam to escape from finite clusters. Suitably chosen fluxes with selected input configurations enables tuning the cage size. Our results have relevance for fundamental properties of topological
lattice and various applications as in non-diffractive image transmission schemes\cite{Vicencio2015,Xia:16}, all-optical logic gates\cite{Real2017} and optical data processing.

\section*{DATA AVAILABILITY}
The data that support the findings of this study are available from the corresponding author upon reasonable request.

\begin{acknowledgments}
Useful discussions with R. Fazio are gratefully acknowledged.

We acknowledge funding from QuantERA ERA-NET Co-fund (Grant
No. 731473, project QUOMPLEX) and H2020 PhoQus project
(Grant No. 820392).
\end{acknowledgments}

\nocite{*}

\bibliography{aip1}

\providecommand{\noopsort}[1]{}\providecommand{\singleletter}[1]{#1}%
\begin{thebibliography}{40}%
\makeatletter
\providecommand \@ifxundefined [1]{%
 \@ifx{#1\undefined}
}%
\providecommand \@ifnum [1]{%
 \ifnum #1\expandafter \@firstoftwo
 \else \expandafter \@secondoftwo
 \fi
}%
\providecommand \@ifx [1]{%
 \ifx #1\expandafter \@firstoftwo
 \else \expandafter \@secondoftwo
 \fi
}%
\providecommand \natexlab [1]{#1}%
\providecommand \enquote  [1]{``#1''}%
\providecommand \bibnamefont  [1]{#1}%
\providecommand \bibfnamefont [1]{#1}%
\providecommand \citenamefont [1]{#1}%
\providecommand \href@noop [0]{\@secondoftwo}%
\providecommand \href [0]{\begingroup \@sanitize@url \@href}%
\providecommand \@href[1]{\@@startlink{#1}\@@href}%
\providecommand \@@href[1]{\endgroup#1\@@endlink}%
\providecommand \@sanitize@url [0]{\catcode `\\12\catcode `\$12\catcode
  `\&12\catcode `\#12\catcode `\^12\catcode `\_12\catcode `\%12\relax}%
\providecommand \@@startlink[1]{}%
\providecommand \@@endlink[0]{}%
\providecommand \url  [0]{\begingroup\@sanitize@url \@url }%
\providecommand \@url [1]{\endgroup\@href {#1}{\urlprefix }}%
\providecommand \urlprefix  [0]{URL }%
\providecommand \Eprint [0]{\href }%
\providecommand \doibase [0]{http://dx.doi.org/}%
\providecommand \selectlanguage [0]{\@gobble}%
\providecommand \bibinfo  [0]{\@secondoftwo}%
\providecommand \bibfield  [0]{\@secondoftwo}%
\providecommand \translation [1]{[#1]}%
\providecommand \BibitemOpen [0]{}%
\providecommand \bibitemStop [0]{}%
\providecommand \bibitemNoStop [0]{.\EOS\space}%
\providecommand \EOS [0]{\spacefactor3000\relax}%
\providecommand \BibitemShut  [1]{\csname bibitem#1\endcsname}%
\let\auto@bib@innerbib\@empty
\bibitem [{\citenamefont {Vanderbilt}(2018)}]{Vanderbilt2018}%
  \BibitemOpen
  \bibfield  {author} {\bibinfo {author} {\bibfnamefont {D.}~\bibnamefont
  {Vanderbilt}},\ }\href {\doibase 10.1017/9781316662205} {\emph {\bibinfo
  {title} {Berry Phases in Electronic Structure Theory}}}\ (\bibinfo
  {publisher} {Cambridge University Press},\ \bibinfo {year}
  {2018})\BibitemShut {NoStop}%
\bibitem [{\citenamefont {Thouless}\ \emph {et~al.}(1982)\citenamefont
  {Thouless}, \citenamefont {Kohmoto}, \citenamefont {Nightingale},\ and\
  \citenamefont {den Nijs}}]{Thouless1982}%
  \BibitemOpen
  \bibfield  {author} {\bibinfo {author} {\bibfnamefont {D.~J.}\ \bibnamefont
  {Thouless}}, \bibinfo {author} {\bibfnamefont {M.}~\bibnamefont {Kohmoto}},
  \bibinfo {author} {\bibfnamefont {M.~P.}\ \bibnamefont {Nightingale}}, \ and\
  \bibinfo {author} {\bibfnamefont {M.}~\bibnamefont {den Nijs}},\ }\bibfield
  {title} {\enquote {\bibinfo {title} {Quantized hall conductance in a
  two-dimensional periodic potential},}\ }\href {\doibase
  10.1103/physrevlett.49.405} {\bibfield  {journal} {\bibinfo  {journal}
  {Physical Review Letters}\ }\textbf {\bibinfo {volume} {49}},\ \bibinfo
  {pages} {405--408} (\bibinfo {year} {1982})}\BibitemShut {NoStop}%
\bibitem [{\citenamefont {Resta}(1993)}]{Resta1993}%
  \BibitemOpen
  \bibfield  {author} {\bibinfo {author} {\bibfnamefont {R.}~\bibnamefont
  {Resta}},\ }\bibfield  {title} {\enquote {\bibinfo {title} {Macroscopic
  electric polarization as a geometric quantum phase},}\ }\href {\doibase
  10.1209/0295-5075/22/2/010} {\bibfield  {journal} {\bibinfo  {journal}
  {Europhysics Letters ({EPL})}\ }\textbf {\bibinfo {volume} {22}},\ \bibinfo
  {pages} {133--138} (\bibinfo {year} {1993})}\BibitemShut {NoStop}%
\bibitem [{\citenamefont {Hasan}\ and\ \citenamefont {Kane}(2010)}]{hasan2010}%
  \BibitemOpen
  \bibfield  {author} {\bibinfo {author} {\bibfnamefont {M.~Z.}\ \bibnamefont
  {Hasan}}\ and\ \bibinfo {author} {\bibfnamefont {C.~L.}\ \bibnamefont
  {Kane}},\ }\bibfield  {title} {\enquote {\bibinfo {title} {Colloquium:
  Topological insulators},}\ }\href {\doibase 10.1103/revmodphys.82.3045}
  {\bibfield  {journal} {\bibinfo  {journal} {Rev. Mod. Phys.}\ }\textbf
  {\bibinfo {volume} {82}},\ \bibinfo {pages} {3045--3067} (\bibinfo {year}
  {2010})}\BibitemShut {NoStop}%
\bibitem [{\citenamefont {Wilczek}\ and\ \citenamefont
  {Shapere}(1989)}]{Wilczek1989}%
  \BibitemOpen
  \bibfield  {author} {\bibinfo {author} {\bibfnamefont {F.}~\bibnamefont
  {Wilczek}}\ and\ \bibinfo {author} {\bibfnamefont {A.}~\bibnamefont
  {Shapere}},\ }\href {\doibase 10.1142/0613} {\emph {\bibinfo {title}
  {Geometric Phases in Physics}}}\ (\bibinfo  {publisher} {{WORLD}
  {SCIENTIFIC}},\ \bibinfo {year} {1989})\BibitemShut {NoStop}%
\bibitem [{\citenamefont {Aidelsburger}, \citenamefont {Nascimbene},\ and\
  \citenamefont {Goldman}(2018)}]{aidelsburger2018}%
  \BibitemOpen
  \bibfield  {author} {\bibinfo {author} {\bibfnamefont {M.}~\bibnamefont
  {Aidelsburger}}, \bibinfo {author} {\bibfnamefont {S.}~\bibnamefont
  {Nascimbene}}, \ and\ \bibinfo {author} {\bibfnamefont {N.}~\bibnamefont
  {Goldman}},\ }\bibfield  {title} {\enquote {\bibinfo {title} {Artificial
  gauge fields in materials and engineered systems},}\ }\href {\doibase
  10.1016/j.crhy.2018.03.002} {\bibfield  {journal} {\bibinfo  {journal}
  {Comptes Rendus Physique}\ }\textbf {\bibinfo {volume} {19}},\ \bibinfo
  {pages} {394--432} (\bibinfo {year} {2018})}\BibitemShut {NoStop}%
\bibitem [{\citenamefont {Aidelsburger}\ \emph {et~al.}(2011)\citenamefont
  {Aidelsburger}, \citenamefont {Atala}, \citenamefont {Nascimb{\`{e}}ne},
  \citenamefont {Trotzky}, \citenamefont {Chen},\ and\ \citenamefont
  {Bloch}}]{Aidelsburger2011}%
  \BibitemOpen
  \bibfield  {author} {\bibinfo {author} {\bibfnamefont {M.}~\bibnamefont
  {Aidelsburger}}, \bibinfo {author} {\bibfnamefont {M.}~\bibnamefont {Atala}},
  \bibinfo {author} {\bibfnamefont {S.}~\bibnamefont {Nascimb{\`{e}}ne}},
  \bibinfo {author} {\bibfnamefont {S.}~\bibnamefont {Trotzky}}, \bibinfo
  {author} {\bibfnamefont {Y.-A.}\ \bibnamefont {Chen}}, \ and\ \bibinfo
  {author} {\bibfnamefont {I.}~\bibnamefont {Bloch}},\ }\bibfield  {title}
  {\enquote {\bibinfo {title} {Experimental realization of strong effective
  magnetic fields in an optical lattice},}\ }\href {\doibase
  10.1103/physrevlett.107.255301} {\bibfield  {journal} {\bibinfo  {journal}
  {Physical Review Letters}\ }\textbf {\bibinfo {volume} {107}},\ \bibinfo
  {pages} {255301} (\bibinfo {year} {2011})}\BibitemShut {NoStop}%
\bibitem [{\citenamefont {Hafezi}\ \emph {et~al.}(2013)\citenamefont {Hafezi},
  \citenamefont {Mittal}, \citenamefont {Fan}, \citenamefont {Migdall},\ and\
  \citenamefont {Taylor}}]{hafezi2013}%
  \BibitemOpen
  \bibfield  {author} {\bibinfo {author} {\bibfnamefont {M.}~\bibnamefont
  {Hafezi}}, \bibinfo {author} {\bibfnamefont {S.}~\bibnamefont {Mittal}},
  \bibinfo {author} {\bibfnamefont {J.}~\bibnamefont {Fan}}, \bibinfo {author}
  {\bibfnamefont {A.}~\bibnamefont {Migdall}}, \ and\ \bibinfo {author}
  {\bibfnamefont {J.~M.}\ \bibnamefont {Taylor}},\ }\bibfield  {title}
  {\enquote {\bibinfo {title} {Imaging topological edge states in silicon
  photonics},}\ }\href {\doibase 10.1038/nphoton.2013.274} {\bibfield
  {journal} {\bibinfo  {journal} {Nature Photonics}\ }\textbf {\bibinfo
  {volume} {7}},\ \bibinfo {pages} {1001--1005} (\bibinfo {year}
  {2013})}\BibitemShut {NoStop}%
\bibitem [{\citenamefont {Schmidt}\ \emph {et~al.}(2015)\citenamefont
  {Schmidt}, \citenamefont {Kessler}, \citenamefont {Peano}, \citenamefont
  {Painter},\ and\ \citenamefont {Marquardt}}]{schmidt2015}%
  \BibitemOpen
  \bibfield  {author} {\bibinfo {author} {\bibfnamefont {M.}~\bibnamefont
  {Schmidt}}, \bibinfo {author} {\bibfnamefont {S.}~\bibnamefont {Kessler}},
  \bibinfo {author} {\bibfnamefont {V.}~\bibnamefont {Peano}}, \bibinfo
  {author} {\bibfnamefont {O.}~\bibnamefont {Painter}}, \ and\ \bibinfo
  {author} {\bibfnamefont {F.}~\bibnamefont {Marquardt}},\ }\bibfield  {title}
  {\enquote {\bibinfo {title} {Optomechanical creation of magnetic fields for
  photons on a lattice},}\ }\href {\doibase 10.1364/optica.2.000635} {\bibfield
   {journal} {\bibinfo  {journal} {Optica}\ }\textbf {\bibinfo {volume} {2}},\
  \bibinfo {pages} {635} (\bibinfo {year} {2015})}\BibitemShut {NoStop}%
\bibitem [{\citenamefont {Rechtsman}\ \emph {et~al.}(2012)\citenamefont
  {Rechtsman}, \citenamefont {Zeuner}, \citenamefont {Tünnermann},
  \citenamefont {Nolte}, \citenamefont {Segev},\ and\ \citenamefont
  {Szameit}}]{rechtsman2012}%
  \BibitemOpen
  \bibfield  {author} {\bibinfo {author} {\bibfnamefont {M.~C.}\ \bibnamefont
  {Rechtsman}}, \bibinfo {author} {\bibfnamefont {J.~M.}\ \bibnamefont
  {Zeuner}}, \bibinfo {author} {\bibfnamefont {A.}~\bibnamefont {Tünnermann}},
  \bibinfo {author} {\bibfnamefont {S.}~\bibnamefont {Nolte}}, \bibinfo
  {author} {\bibfnamefont {M.}~\bibnamefont {Segev}}, \ and\ \bibinfo {author}
  {\bibfnamefont {A.}~\bibnamefont {Szameit}},\ }\bibfield  {title} {\enquote
  {\bibinfo {title} {Strain-induced pseudomagnetic field and photonic landau
  levels in dielectric structures},}\ }\href {\doibase
  10.1038/nphoton.2012.302} {\bibfield  {journal} {\bibinfo  {journal} {Nature
  Photonics}\ }\textbf {\bibinfo {volume} {7}},\ \bibinfo {pages} {153--158}
  (\bibinfo {year} {2012})}\BibitemShut {NoStop}%
\bibitem [{\citenamefont {Lumer}\ \emph {et~al.}(2019)\citenamefont {Lumer},
  \citenamefont {Bandres}, \citenamefont {Heinrich}, \citenamefont {Maczewsky},
  \citenamefont {Herzig-Sheinfux}, \citenamefont {Szameit},\ and\ \citenamefont
  {Segev}}]{lumer2019}%
  \BibitemOpen
  \bibfield  {author} {\bibinfo {author} {\bibfnamefont {Y.}~\bibnamefont
  {Lumer}}, \bibinfo {author} {\bibfnamefont {M.~A.}\ \bibnamefont {Bandres}},
  \bibinfo {author} {\bibfnamefont {M.}~\bibnamefont {Heinrich}}, \bibinfo
  {author} {\bibfnamefont {L.~J.}\ \bibnamefont {Maczewsky}}, \bibinfo {author}
  {\bibfnamefont {H.}~\bibnamefont {Herzig-Sheinfux}}, \bibinfo {author}
  {\bibfnamefont {A.}~\bibnamefont {Szameit}}, \ and\ \bibinfo {author}
  {\bibfnamefont {M.}~\bibnamefont {Segev}},\ }\bibfield  {title} {\enquote
  {\bibinfo {title} {Light guiding by artificial gauge fields},}\ }\href
  {\doibase 10.1038/s41566-019-0370-1} {\bibfield  {journal} {\bibinfo
  {journal} {Nature Photonics}\ }\textbf {\bibinfo {volume} {13}},\ \bibinfo
  {pages} {339--345} (\bibinfo {year} {2019})}\BibitemShut {NoStop}%
\bibitem [{\citenamefont {Fang}, \citenamefont {Yu},\ and\ \citenamefont
  {Fan}(2012{\natexlab{a}})}]{fang2012}%
  \BibitemOpen
  \bibfield  {author} {\bibinfo {author} {\bibfnamefont {K.}~\bibnamefont
  {Fang}}, \bibinfo {author} {\bibfnamefont {Z.}~\bibnamefont {Yu}}, \ and\
  \bibinfo {author} {\bibfnamefont {S.}~\bibnamefont {Fan}},\ }\bibfield
  {title} {\enquote {\bibinfo {title} {Realizing effective magnetic field for
  photons by controlling the phase of dynamic modulation},}\ }\href {\doibase
  10.1038/nphoton.2012.236} {\bibfield  {journal} {\bibinfo  {journal} {Nature
  Photonics}\ }\textbf {\bibinfo {volume} {6}},\ \bibinfo {pages} {782--787}
  (\bibinfo {year} {2012}{\natexlab{a}})}\BibitemShut {NoStop}%
\bibitem [{\citenamefont {Goldman}\ and\ \citenamefont
  {Dalibard}(2014)}]{goldman2014}%
  \BibitemOpen
  \bibfield  {author} {\bibinfo {author} {\bibfnamefont {N.}~\bibnamefont
  {Goldman}}\ and\ \bibinfo {author} {\bibfnamefont {J.}~\bibnamefont
  {Dalibard}},\ }\bibfield  {title} {\enquote {\bibinfo {title} {Periodically
  driven quantum systems: Effective hamiltonians and engineered gauge
  fields},}\ }\href {\doibase 10.1103/physrevx.4.031027} {\bibfield  {journal}
  {\bibinfo  {journal} {Physical Review X}\ }\textbf {\bibinfo {volume} {4}}
  (\bibinfo {year} {2014}),\ 10.1103/physrevx.4.031027}\BibitemShut {NoStop}%
\bibitem [{\citenamefont {J\"{o}rg}\ \emph {et~al.}(2017)\citenamefont
  {J\"{o}rg}, \citenamefont {Letscher}, \citenamefont {Fleischhauer},\ and\
  \citenamefont {von Freymann}}]{joerg2017}%
  \BibitemOpen
  \bibfield  {author} {\bibinfo {author} {\bibfnamefont {C.}~\bibnamefont
  {J\"{o}rg}}, \bibinfo {author} {\bibfnamefont {F.}~\bibnamefont {Letscher}},
  \bibinfo {author} {\bibfnamefont {M.}~\bibnamefont {Fleischhauer}}, \ and\
  \bibinfo {author} {\bibfnamefont {G.}~\bibnamefont {von Freymann}},\
  }\bibfield  {title} {\enquote {\bibinfo {title} {Dynamic defects in photonic
  floquet topological insulators},}\ }\href {\doibase 10.1088/1367-2630/aa7c82}
  {\bibfield  {journal} {\bibinfo  {journal} {New Journal of Physics}\ }\textbf
  {\bibinfo {volume} {19}},\ \bibinfo {pages} {083003} (\bibinfo {year}
  {2017})}\BibitemShut {NoStop}%
\bibitem [{\citenamefont {Lustig}\ \emph {et~al.}(2019)\citenamefont {Lustig},
  \citenamefont {Weimann}, \citenamefont {Plotnik}, \citenamefont {Lumer},
  \citenamefont {Bandres}, \citenamefont {Szameit},\ and\ \citenamefont
  {Segev}}]{lustig2019}%
  \BibitemOpen
  \bibfield  {author} {\bibinfo {author} {\bibfnamefont {E.}~\bibnamefont
  {Lustig}}, \bibinfo {author} {\bibfnamefont {S.}~\bibnamefont {Weimann}},
  \bibinfo {author} {\bibfnamefont {Y.}~\bibnamefont {Plotnik}}, \bibinfo
  {author} {\bibfnamefont {Y.}~\bibnamefont {Lumer}}, \bibinfo {author}
  {\bibfnamefont {M.~A.}\ \bibnamefont {Bandres}}, \bibinfo {author}
  {\bibfnamefont {A.}~\bibnamefont {Szameit}}, \ and\ \bibinfo {author}
  {\bibfnamefont {M.}~\bibnamefont {Segev}},\ }\bibfield  {title} {\enquote
  {\bibinfo {title} {Photonic topological insulator in synthetic dimensions},}\
  }\href {\doibase 10.1038/s41586-019-0943-7} {\bibfield  {journal} {\bibinfo
  {journal} {Nature}\ }\textbf {\bibinfo {volume} {567}},\ \bibinfo {pages}
  {356--360} (\bibinfo {year} {2019})}\BibitemShut {NoStop}%
\bibitem [{\citenamefont {J\"{o}rg}\ \emph {et~al.}(2020)\citenamefont
  {J\"{o}rg}, \citenamefont {Queralt{\'{o}}}, \citenamefont {Kremer},
  \citenamefont {Pelegr{\'{\i}}}, \citenamefont {Schulz}, \citenamefont
  {Szameit}, \citenamefont {von Freymann}, \citenamefont {Mompart},\ and\
  \citenamefont {Ahufinger}}]{joerg2020}%
  \BibitemOpen
  \bibfield  {author} {\bibinfo {author} {\bibfnamefont {C.}~\bibnamefont
  {J\"{o}rg}}, \bibinfo {author} {\bibfnamefont {G.}~\bibnamefont
  {Queralt{\'{o}}}}, \bibinfo {author} {\bibfnamefont {M.}~\bibnamefont
  {Kremer}}, \bibinfo {author} {\bibfnamefont {G.}~\bibnamefont
  {Pelegr{\'{\i}}}}, \bibinfo {author} {\bibfnamefont {J.}~\bibnamefont
  {Schulz}}, \bibinfo {author} {\bibfnamefont {A.}~\bibnamefont {Szameit}},
  \bibinfo {author} {\bibfnamefont {G.}~\bibnamefont {von Freymann}}, \bibinfo
  {author} {\bibfnamefont {J.}~\bibnamefont {Mompart}}, \ and\ \bibinfo
  {author} {\bibfnamefont {V.}~\bibnamefont {Ahufinger}},\ }\bibfield  {title}
  {\enquote {\bibinfo {title} {Artificial gauge field switching using orbital
  angular momentum modes in optical waveguides},}\ }\href {\doibase
  10.1038/s41377-020-00385-6} {\bibfield  {journal} {\bibinfo  {journal}
  {Light: Science {\&} Applications}\ }\textbf {\bibinfo {volume} {9}}
  (\bibinfo {year} {2020}),\ 10.1038/s41377-020-00385-6}\BibitemShut {NoStop}%
\bibitem [{\citenamefont {Aharonov}\ and\ \citenamefont
  {Bohm}(1959)}]{aharonov1959}%
  \BibitemOpen
  \bibfield  {author} {\bibinfo {author} {\bibfnamefont {Y.}~\bibnamefont
  {Aharonov}}\ and\ \bibinfo {author} {\bibfnamefont {D.}~\bibnamefont
  {Bohm}},\ }\bibfield  {title} {\enquote {\bibinfo {title} {Significance of
  electromagnetic potentials in the quantum theory},}\ }\href {\doibase
  10.1103/physrev.115.485} {\bibfield  {journal} {\bibinfo  {journal} {Phys.
  Rev.}\ }\textbf {\bibinfo {volume} {115}},\ \bibinfo {pages} {485--491}
  (\bibinfo {year} {1959})}\BibitemShut {NoStop}%
\bibitem [{\citenamefont {Batelaan}\ and\ \citenamefont
  {Tonomura}(2009)}]{batelaan2009}%
  \BibitemOpen
  \bibfield  {author} {\bibinfo {author} {\bibfnamefont {H.}~\bibnamefont
  {Batelaan}}\ and\ \bibinfo {author} {\bibfnamefont {A.}~\bibnamefont
  {Tonomura}},\ }\bibfield  {title} {\enquote {\bibinfo {title} {The
  aharonov{\textendash}bohm effects: Variations on a subtle theme},}\ }\href
  {\doibase 10.1063/1.3226854} {\bibfield  {journal} {\bibinfo  {journal}
  {Physics Today}\ }\textbf {\bibinfo {volume} {62}},\ \bibinfo {pages}
  {38--43} (\bibinfo {year} {2009})}\BibitemShut {NoStop}%
\bibitem [{\citenamefont {Wu}\ and\ \citenamefont {Yang}(1975)}]{wu1975}%
  \BibitemOpen
  \bibfield  {author} {\bibinfo {author} {\bibfnamefont {T.~T.}\ \bibnamefont
  {Wu}}\ and\ \bibinfo {author} {\bibfnamefont {C.~N.}\ \bibnamefont {Yang}},\
  }\bibfield  {title} {\enquote {\bibinfo {title} {Concept of nonintegrable
  phase factors and global formulation of gauge fields},}\ }\href {\doibase
  10.1103/physrevd.12.3845} {\bibfield  {journal} {\bibinfo  {journal}
  {Physical Review D}\ }\textbf {\bibinfo {volume} {12}},\ \bibinfo {pages}
  {3845--3857} (\bibinfo {year} {1975})}\BibitemShut {NoStop}%
\bibitem [{\citenamefont {Vidal}, \citenamefont {Mosseri},\ and\ \citenamefont
  {Dou{\c{c}}ot}(1998)}]{vidal1998}%
  \BibitemOpen
  \bibfield  {author} {\bibinfo {author} {\bibfnamefont {J.}~\bibnamefont
  {Vidal}}, \bibinfo {author} {\bibfnamefont {R.}~\bibnamefont {Mosseri}}, \
  and\ \bibinfo {author} {\bibfnamefont {B.}~\bibnamefont {Dou{\c{c}}ot}},\
  }\bibfield  {title} {\enquote {\bibinfo {title} {Aharonov-bohm cages in
  two-dimensional structures},}\ }\href {\doibase 10.1103/physrevlett.81.5888}
  {\bibfield  {journal} {\bibinfo  {journal} {Physical Review Letters}\
  }\textbf {\bibinfo {volume} {81}},\ \bibinfo {pages} {5888--5891} (\bibinfo
  {year} {1998})}\BibitemShut {NoStop}%
\bibitem [{\citenamefont {Leykam}, \citenamefont {Andreanov},\ and\
  \citenamefont {Flach}(2018)}]{leykam2018}%
  \BibitemOpen
  \bibfield  {author} {\bibinfo {author} {\bibfnamefont {D.}~\bibnamefont
  {Leykam}}, \bibinfo {author} {\bibfnamefont {A.}~\bibnamefont {Andreanov}}, \
  and\ \bibinfo {author} {\bibfnamefont {S.}~\bibnamefont {Flach}},\ }\bibfield
   {title} {\enquote {\bibinfo {title} {Artificial flat band systems: from
  lattice models to experiments},}\ }\href {\doibase
  10.1080/23746149.2018.1473052} {\bibfield  {journal} {\bibinfo  {journal}
  {Advances in Physics: X}\ }\textbf {\bibinfo {volume} {3}},\ \bibinfo {pages}
  {1473052} (\bibinfo {year} {2018})}\BibitemShut {NoStop}%
\bibitem [{\citenamefont {Abilio}\ \emph {et~al.}(1999)\citenamefont {Abilio},
  \citenamefont {Butaud}, \citenamefont {Fournier}, \citenamefont {Pannetier},
  \citenamefont {Vidal}, \citenamefont {Tedesco},\ and\ \citenamefont
  {Dalzotto}}]{abilio1999}%
  \BibitemOpen
  \bibfield  {author} {\bibinfo {author} {\bibfnamefont {C.~C.}\ \bibnamefont
  {Abilio}}, \bibinfo {author} {\bibfnamefont {P.}~\bibnamefont {Butaud}},
  \bibinfo {author} {\bibfnamefont {T.}~\bibnamefont {Fournier}}, \bibinfo
  {author} {\bibfnamefont {B.}~\bibnamefont {Pannetier}}, \bibinfo {author}
  {\bibfnamefont {J.}~\bibnamefont {Vidal}}, \bibinfo {author} {\bibfnamefont
  {S.}~\bibnamefont {Tedesco}}, \ and\ \bibinfo {author} {\bibfnamefont
  {B.}~\bibnamefont {Dalzotto}},\ }\bibfield  {title} {\enquote {\bibinfo
  {title} {Magnetic field induced localization in a two-dimensional
  superconducting wire network},}\ }\href {\doibase
  10.1103/physrevlett.83.5102} {\bibfield  {journal} {\bibinfo  {journal}
  {Physical Review Letters}\ }\textbf {\bibinfo {volume} {83}},\ \bibinfo
  {pages} {5102--5105} (\bibinfo {year} {1999})}\BibitemShut {NoStop}%
\bibitem [{\citenamefont {Rizzi}, \citenamefont {Cataudella},\ and\
  \citenamefont {Fazio}(2006)}]{rizzi2006}%
  \BibitemOpen
  \bibfield  {author} {\bibinfo {author} {\bibfnamefont {M.}~\bibnamefont
  {Rizzi}}, \bibinfo {author} {\bibfnamefont {V.}~\bibnamefont {Cataudella}}, \
  and\ \bibinfo {author} {\bibfnamefont {R.}~\bibnamefont {Fazio}},\ }\bibfield
   {title} {\enquote {\bibinfo {title} {Phase diagram of the bose-hubbard model
  {withT}3symmetry},}\ }\href {\doibase 10.1103/physrevb.73.144511} {\bibfield
  {journal} {\bibinfo  {journal} {Physical Review B}\ }\textbf {\bibinfo
  {volume} {73}} (\bibinfo {year} {2006}),\
  10.1103/physrevb.73.144511}\BibitemShut {NoStop}%
\bibitem [{\citenamefont {Mukherjee}\ \emph {et~al.}(2018)\citenamefont
  {Mukherjee}, \citenamefont {Liberto}, \citenamefont {Öhberg}, \citenamefont
  {Thomson},\ and\ \citenamefont {Goldman}}]{mukherjee2018}%
  \BibitemOpen
  \bibfield  {author} {\bibinfo {author} {\bibfnamefont {S.}~\bibnamefont
  {Mukherjee}}, \bibinfo {author} {\bibfnamefont {M.~D.}\ \bibnamefont
  {Liberto}}, \bibinfo {author} {\bibfnamefont {P.}~\bibnamefont {Öhberg}},
  \bibinfo {author} {\bibfnamefont {R.~R.}\ \bibnamefont {Thomson}}, \ and\
  \bibinfo {author} {\bibfnamefont {N.}~\bibnamefont {Goldman}},\ }\bibfield
  {title} {\enquote {\bibinfo {title} {Experimental observation of
  aharonov-bohm cages in photonic lattices},}\ }\href {\doibase
  10.1103/physrevlett.121.075502} {\bibfield  {journal} {\bibinfo  {journal}
  {Physical Review Letters}\ }\textbf {\bibinfo {volume} {121}} (\bibinfo
  {year} {2018}),\ 10.1103/physrevlett.121.075502}\BibitemShut {NoStop}%
\bibitem [{\citenamefont {Kremer}\ \emph {et~al.}(2020)\citenamefont {Kremer},
  \citenamefont {Petrides}, \citenamefont {Meyer}, \citenamefont {Heinrich},
  \citenamefont {Zilberberg},\ and\ \citenamefont {Szameit}}]{kremer2020}%
  \BibitemOpen
  \bibfield  {author} {\bibinfo {author} {\bibfnamefont {M.}~\bibnamefont
  {Kremer}}, \bibinfo {author} {\bibfnamefont {I.}~\bibnamefont {Petrides}},
  \bibinfo {author} {\bibfnamefont {E.}~\bibnamefont {Meyer}}, \bibinfo
  {author} {\bibfnamefont {M.}~\bibnamefont {Heinrich}}, \bibinfo {author}
  {\bibfnamefont {O.}~\bibnamefont {Zilberberg}}, \ and\ \bibinfo {author}
  {\bibfnamefont {A.}~\bibnamefont {Szameit}},\ }\bibfield  {title} {\enquote
  {\bibinfo {title} {A square-root topological insulator with non-quantized
  indices realized with photonic aharonov-bohm cages},}\ }\href {\doibase
  10.1038/s41467-020-14692-4} {\bibfield  {journal} {\bibinfo  {journal}
  {Nature Communications}\ }\textbf {\bibinfo {volume} {11}} (\bibinfo {year}
  {2020}),\ 10.1038/s41467-020-14692-4}\BibitemShut {NoStop}%
\bibitem [{\citenamefont {Brosco}\ \emph {et~al.}(2020)\citenamefont {Brosco},
  \citenamefont {Pilozzi}, \citenamefont {Fazio},\ and\ \citenamefont
  {Conti}}]{brosco2020}%
  \BibitemOpen
  \bibfield  {author} {\bibinfo {author} {\bibfnamefont {V.}~\bibnamefont
  {Brosco}}, \bibinfo {author} {\bibfnamefont {L.}~\bibnamefont {Pilozzi}},
  \bibinfo {author} {\bibfnamefont {R.}~\bibnamefont {Fazio}}, \ and\ \bibinfo
  {author} {\bibfnamefont {C.}~\bibnamefont {Conti}},\ }\href@noop {} {\enquote
  {\bibinfo {title} {Non-abelian thouless pumping in a photonic lattice},}\ }
  (\bibinfo {year} {2020}),\ \Eprint {http://arxiv.org/abs/2010.15159}
  {arXiv:2010.15159 [cond-mat.mes-hall]} \BibitemShut {NoStop}%
\bibitem [{\citenamefont {Longhi}(2014)}]{Longhi2014}%
  \BibitemOpen
  \bibfield  {author} {\bibinfo {author} {\bibfnamefont {S.}~\bibnamefont
  {Longhi}},\ }\bibfield  {title} {\enquote {\bibinfo {title}
  {Aharonov{\textendash}bohm photonic cages in waveguide and coupled resonator
  lattices by synthetic magnetic fields},}\ }\href {\doibase
  10.1364/ol.39.005892} {\bibfield  {journal} {\bibinfo  {journal} {Optics
  Letters}\ }\textbf {\bibinfo {volume} {39}},\ \bibinfo {pages} {5892}
  (\bibinfo {year} {2014})}\BibitemShut {NoStop}%
\bibitem [{\citenamefont {Christodoulides}, \citenamefont {Lederer},\ and\
  \citenamefont {Silberberg}(2003)}]{christodoulides2003}%
  \BibitemOpen
  \bibfield  {author} {\bibinfo {author} {\bibfnamefont {D.~N.}\ \bibnamefont
  {Christodoulides}}, \bibinfo {author} {\bibfnamefont {F.}~\bibnamefont
  {Lederer}}, \ and\ \bibinfo {author} {\bibfnamefont {Y.}~\bibnamefont
  {Silberberg}},\ }\bibfield  {title} {\enquote {\bibinfo {title} {Discretizing
  light behaviour in linear and nonlinear waveguide lattices},}\ }\href
  {\doibase 10.1038/nature01936} {\bibfield  {journal} {\bibinfo  {journal}
  {Nature}\ }\textbf {\bibinfo {volume} {424}},\ \bibinfo {pages} {817--823}
  (\bibinfo {year} {2003})}\BibitemShut {NoStop}%
\bibitem [{\citenamefont {Vicencio}\ \emph {et~al.}(2015)\citenamefont
  {Vicencio}, \citenamefont {Cantillano}, \citenamefont {Morales-Inostroza},
  \citenamefont {Real}, \citenamefont {Mej{\'{\i}}a-Cort{\'{e}}s},
  \citenamefont {Weimann}, \citenamefont {Szameit},\ and\ \citenamefont
  {Molina}}]{Vicencio2015}%
  \BibitemOpen
  \bibfield  {author} {\bibinfo {author} {\bibfnamefont {R.~A.}\ \bibnamefont
  {Vicencio}}, \bibinfo {author} {\bibfnamefont {C.}~\bibnamefont
  {Cantillano}}, \bibinfo {author} {\bibfnamefont {L.}~\bibnamefont
  {Morales-Inostroza}}, \bibinfo {author} {\bibfnamefont {B.}~\bibnamefont
  {Real}}, \bibinfo {author} {\bibfnamefont {C.}~\bibnamefont
  {Mej{\'{\i}}a-Cort{\'{e}}s}}, \bibinfo {author} {\bibfnamefont
  {S.}~\bibnamefont {Weimann}}, \bibinfo {author} {\bibfnamefont
  {A.}~\bibnamefont {Szameit}}, \ and\ \bibinfo {author} {\bibfnamefont
  {M.~I.}\ \bibnamefont {Molina}},\ }\bibfield  {title} {\enquote {\bibinfo
  {title} {Observation of localized states in lieb photonic lattices},}\ }\href
  {\doibase 10.1103/physrevlett.114.245503} {\bibfield  {journal} {\bibinfo
  {journal} {Physical Review Letters}\ }\textbf {\bibinfo {volume} {114}},\
  \bibinfo {pages} {245503} (\bibinfo {year} {2015})}\BibitemShut {NoStop}%
\bibitem [{\citenamefont {Xia}\ \emph {et~al.}(2016)\citenamefont {Xia},
  \citenamefont {Hu}, \citenamefont {Song}, \citenamefont {Zong}, \citenamefont
  {Tang},\ and\ \citenamefont {Chen}}]{Xia:16}%
  \BibitemOpen
  \bibfield  {author} {\bibinfo {author} {\bibfnamefont {S.}~\bibnamefont
  {Xia}}, \bibinfo {author} {\bibfnamefont {Y.}~\bibnamefont {Hu}}, \bibinfo
  {author} {\bibfnamefont {D.}~\bibnamefont {Song}}, \bibinfo {author}
  {\bibfnamefont {Y.}~\bibnamefont {Zong}}, \bibinfo {author} {\bibfnamefont
  {L.}~\bibnamefont {Tang}}, \ and\ \bibinfo {author} {\bibfnamefont
  {Z.}~\bibnamefont {Chen}},\ }\bibfield  {title} {\enquote {\bibinfo {title}
  {Demonstration of flat-band image transmission in optically induced lieb
  photonic lattices},}\ }\href {\doibase 10.1364/OL.41.001435} {\bibfield
  {journal} {\bibinfo  {journal} {Opt. Lett.}\ }\textbf {\bibinfo {volume}
  {41}},\ \bibinfo {pages} {1435--1438} (\bibinfo {year} {2016})}\BibitemShut
  {NoStop}%
\bibitem [{\citenamefont {Real}\ \emph {et~al.}(2017)\citenamefont {Real},
  \citenamefont {Cantillano}, \citenamefont {L{\'{o}}pez-Gonz{\'{a}}lez},
  \citenamefont {Szameit}, \citenamefont {Aono}, \citenamefont {Naruse},
  \citenamefont {Kim}, \citenamefont {Wang},\ and\ \citenamefont
  {Vicencio}}]{Real2017}%
  \BibitemOpen
  \bibfield  {author} {\bibinfo {author} {\bibfnamefont {B.}~\bibnamefont
  {Real}}, \bibinfo {author} {\bibfnamefont {C.}~\bibnamefont {Cantillano}},
  \bibinfo {author} {\bibfnamefont {D.}~\bibnamefont
  {L{\'{o}}pez-Gonz{\'{a}}lez}}, \bibinfo {author} {\bibfnamefont
  {A.}~\bibnamefont {Szameit}}, \bibinfo {author} {\bibfnamefont
  {M.}~\bibnamefont {Aono}}, \bibinfo {author} {\bibfnamefont {M.}~\bibnamefont
  {Naruse}}, \bibinfo {author} {\bibfnamefont {S.-J.}\ \bibnamefont {Kim}},
  \bibinfo {author} {\bibfnamefont {K.}~\bibnamefont {Wang}}, \ and\ \bibinfo
  {author} {\bibfnamefont {R.~A.}\ \bibnamefont {Vicencio}},\ }\bibfield
  {title} {\enquote {\bibinfo {title} {Flat-band light dynamics in stub
  photonic lattices},}\ }\href {\doibase 10.1038/s41598-017-15441-2} {\bibfield
   {journal} {\bibinfo  {journal} {Scientific Reports}\ }\textbf {\bibinfo
  {volume} {7}} (\bibinfo {year} {2017}),\
  10.1038/s41598-017-15441-2}\BibitemShut {NoStop}%
\bibitem [{\citenamefont {Hofstadter}(1976)}]{Hofstadter1976}%
  \BibitemOpen
  \bibfield  {author} {\bibinfo {author} {\bibfnamefont {D.~R.}\ \bibnamefont
  {Hofstadter}},\ }\bibfield  {title} {\enquote {\bibinfo {title} {Energy
  levels and wave functions of bloch electrons in rational and irrational
  magnetic fields},}\ }\href {\doibase 10.1103/physrevb.14.2239} {\bibfield
  {journal} {\bibinfo  {journal} {Physical Review B}\ }\textbf {\bibinfo
  {volume} {14}},\ \bibinfo {pages} {2239--2249} (\bibinfo {year}
  {1976})}\BibitemShut {NoStop}%
\bibitem [{\citenamefont {Langbein}(1969)}]{Langbein1969}%
  \BibitemOpen
  \bibfield  {author} {\bibinfo {author} {\bibfnamefont {D.}~\bibnamefont
  {Langbein}},\ }\bibfield  {title} {\enquote {\bibinfo {title} {The
  tight-binding and the nearly-free-electron approach to lattice electrons in
  external magnetic fields},}\ }\href {\doibase 10.1103/physrev.180.633}
  {\bibfield  {journal} {\bibinfo  {journal} {Physical Review}\ }\textbf
  {\bibinfo {volume} {180}},\ \bibinfo {pages} {633--648} (\bibinfo {year}
  {1969})}\BibitemShut {NoStop}%
\bibitem [{\citenamefont {Hügel}\ and\ \citenamefont
  {Paredes}(2014)}]{Huegel2014}%
  \BibitemOpen
  \bibfield  {author} {\bibinfo {author} {\bibfnamefont {D.}~\bibnamefont
  {Hügel}}\ and\ \bibinfo {author} {\bibfnamefont {B.}~\bibnamefont
  {Paredes}},\ }\bibfield  {title} {\enquote {\bibinfo {title} {Chiral ladders
  and the edges of quantum hall insulators},}\ }\href {\doibase
  10.1103/physreva.89.023619} {\bibfield  {journal} {\bibinfo  {journal}
  {Physical Review A}\ }\textbf {\bibinfo {volume} {89}},\ \bibinfo {pages}
  {023619} (\bibinfo {year} {2014})}\BibitemShut {NoStop}%
\bibitem [{\citenamefont {Ozawa}\ \emph {et~al.}(2019)\citenamefont {Ozawa},
  \citenamefont {Price}, \citenamefont {Amo}, \citenamefont {Goldman},
  \citenamefont {Hafezi}, \citenamefont {Lu}, \citenamefont {Rechtsman},
  \citenamefont {Schuster}, \citenamefont {Simon}, \citenamefont {Zilberberg},\
  and\ \citenamefont {Carusotto}}]{ozawa2019}%
  \BibitemOpen
  \bibfield  {author} {\bibinfo {author} {\bibfnamefont {T.}~\bibnamefont
  {Ozawa}}, \bibinfo {author} {\bibfnamefont {H.~M.}\ \bibnamefont {Price}},
  \bibinfo {author} {\bibfnamefont {A.}~\bibnamefont {Amo}}, \bibinfo {author}
  {\bibfnamefont {N.}~\bibnamefont {Goldman}}, \bibinfo {author} {\bibfnamefont
  {M.}~\bibnamefont {Hafezi}}, \bibinfo {author} {\bibfnamefont
  {L.}~\bibnamefont {Lu}}, \bibinfo {author} {\bibfnamefont {M.~C.}\
  \bibnamefont {Rechtsman}}, \bibinfo {author} {\bibfnamefont {D.}~\bibnamefont
  {Schuster}}, \bibinfo {author} {\bibfnamefont {J.}~\bibnamefont {Simon}},
  \bibinfo {author} {\bibfnamefont {O.}~\bibnamefont {Zilberberg}}, \ and\
  \bibinfo {author} {\bibfnamefont {I.}~\bibnamefont {Carusotto}},\ }\bibfield
  {title} {\enquote {\bibinfo {title} {Topological photonics},}\ }\href
  {\doibase 10.1103/revmodphys.91.015006} {\bibfield  {journal} {\bibinfo
  {journal} {Rev. Mod. Phys.}\ }\textbf {\bibinfo {volume} {91}},\ \bibinfo
  {pages} {015006} (\bibinfo {year} {2019})}\BibitemShut {NoStop}%
\bibitem [{\citenamefont {Ozawa}\ and\ \citenamefont
  {Price}(2019)}]{ozawa2019a}%
  \BibitemOpen
  \bibfield  {author} {\bibinfo {author} {\bibfnamefont {T.}~\bibnamefont
  {Ozawa}}\ and\ \bibinfo {author} {\bibfnamefont {H.~M.}\ \bibnamefont
  {Price}},\ }\bibfield  {title} {\enquote {\bibinfo {title} {Topological
  quantum matter in synthetic dimensions},}\ }\href {\doibase
  10.1038/s42254-019-0045-3} {\bibfield  {journal} {\bibinfo  {journal} {Nature
  Reviews Physics}\ }\textbf {\bibinfo {volume} {1}},\ \bibinfo {pages}
  {349--357} (\bibinfo {year} {2019})}\BibitemShut {NoStop}%
\bibitem [{\citenamefont {Cohen}\ \emph {et~al.}(2019)\citenamefont {Cohen},
  \citenamefont {Larocque}, \citenamefont {Bouchard}, \citenamefont
  {Nejadsattari}, \citenamefont {Gefen},\ and\ \citenamefont
  {Karimi}}]{Cohen2019}%
  \BibitemOpen
  \bibfield  {author} {\bibinfo {author} {\bibfnamefont {E.}~\bibnamefont
  {Cohen}}, \bibinfo {author} {\bibfnamefont {H.}~\bibnamefont {Larocque}},
  \bibinfo {author} {\bibfnamefont {F.}~\bibnamefont {Bouchard}}, \bibinfo
  {author} {\bibfnamefont {F.}~\bibnamefont {Nejadsattari}}, \bibinfo {author}
  {\bibfnamefont {Y.}~\bibnamefont {Gefen}}, \ and\ \bibinfo {author}
  {\bibfnamefont {E.}~\bibnamefont {Karimi}},\ }\bibfield  {title} {\enquote
  {\bibinfo {title} {Geometric phase from aharonov{\textendash}bohm to
  pancharatnam{\textendash}berry and~beyond},}\ }\href {\doibase
  10.1038/s42254-019-0071-1} {\bibfield  {journal} {\bibinfo  {journal} {Nature
  Reviews Physics}\ }\textbf {\bibinfo {volume} {1}},\ \bibinfo {pages}
  {437--449} (\bibinfo {year} {2019})}\BibitemShut {NoStop}%
\bibitem [{\citenamefont {Fang}, \citenamefont {Yu},\ and\ \citenamefont
  {Fan}(2012{\natexlab{b}})}]{fang2012a}%
  \BibitemOpen
  \bibfield  {author} {\bibinfo {author} {\bibfnamefont {K.}~\bibnamefont
  {Fang}}, \bibinfo {author} {\bibfnamefont {Z.}~\bibnamefont {Yu}}, \ and\
  \bibinfo {author} {\bibfnamefont {S.}~\bibnamefont {Fan}},\ }\bibfield
  {title} {\enquote {\bibinfo {title} {Photonic aharonov-bohm effect based on
  dynamic modulation},}\ }\href {\doibase 10.1103/physrevlett.108.153901}
  {\bibfield  {journal} {\bibinfo  {journal} {Physical Review Letters}\
  }\textbf {\bibinfo {volume} {108}} (\bibinfo {year} {2012}{\natexlab{b}}),\
  10.1103/physrevlett.108.153901}\BibitemShut {NoStop}%
\bibitem [{\citenamefont {Vidal}\ \emph {et~al.}(2000)\citenamefont {Vidal},
  \citenamefont {Dou{\c{c}}ot}, \citenamefont {Mosseri},\ and\ \citenamefont
  {Butaud}}]{vidal2000}%
  \BibitemOpen
  \bibfield  {author} {\bibinfo {author} {\bibfnamefont {J.}~\bibnamefont
  {Vidal}}, \bibinfo {author} {\bibfnamefont {B.}~\bibnamefont {Dou{\c{c}}ot}},
  \bibinfo {author} {\bibfnamefont {R.}~\bibnamefont {Mosseri}}, \ and\
  \bibinfo {author} {\bibfnamefont {P.}~\bibnamefont {Butaud}},\ }\bibfield
  {title} {\enquote {\bibinfo {title} {Interaction induced delocalization for
  two particles in a periodic potential},}\ }\href {\doibase
  10.1103/physrevlett.85.3906} {\bibfield  {journal} {\bibinfo  {journal}
  {Physical Review Letters}\ }\textbf {\bibinfo {volume} {85}},\ \bibinfo
  {pages} {3906--3909} (\bibinfo {year} {2000})}\BibitemShut {NoStop}%
\bibitem [{\citenamefont {Dou{\c{c}}ot}\ and\ \citenamefont
  {Vidal}(2002)}]{doucot2002}%
  \BibitemOpen
  \bibfield  {author} {\bibinfo {author} {\bibfnamefont {B.}~\bibnamefont
  {Dou{\c{c}}ot}}\ and\ \bibinfo {author} {\bibfnamefont {J.}~\bibnamefont
  {Vidal}},\ }\bibfield  {title} {\enquote {\bibinfo {title} {Pairing of cooper
  pairs in a fully frustrated josephson-junction chain},}\ }\href {\doibase
  10.1103/physrevlett.88.227005} {\bibfield  {journal} {\bibinfo  {journal}
  {Physical Review Letters}\ }\textbf {\bibinfo {volume} {88}} (\bibinfo {year}
  {2002}),\ 10.1103/physrevlett.88.227005}\BibitemShut {NoStop}%
\end{thebibliography}%

\end{document}